\documentclass[preprint,11pt]{elsarticle}

\usepackage{amssymb}
\usepackage{amsmath}
\usepackage{epsfig}
\usepackage{subfigure}
\usepackage{url}
\usepackage{multirow}

\newenvironment{list2}{
  \begin{list}{$\bullet$}{%
      \setlength{\itemsep}{0in}
      \setlength{\parsep}{0in} \setlength{\parskip}{0in}
      \setlength{\topsep}{0in} \setlength{\partopsep}{0in} 
      \setlength{\leftmargin}{0.5in}}}{\end{list}}

\journal{arxiv.org}

\begin{document}

\begin{frontmatter}

\title{GPU-accelerated Parallel Solutions to the Quadratic Assignment Problem}


\author{Clara Novoa}
\ead{cn17@txstate.edu}
\address{Ingram School of Engineering\\Texas State University}

\author{Apan Qasem\corref{cor1}}
\ead{apan@txstate.edu}
\address{Dept. of Computer Science\\Texas State University}


\begin{abstract}
The Quadratic Assignment Problem (QAP) is an important combinatorial optimization problem with applications  
in many areas including logistics and manufacturing.
QAP is known to be NP-hard, a computationally challenging problem, which requires the use of
sophisticated heuristics in finding acceptable solutions for most real-world data sets.


In this paper, we present GPU-accelerated implementations of 
a \textit{2opt} and a \textit{tabu search} algorithm for solving the
QAP. For both algorithms, we extract parallelism at multiple levels and implement novel code
optimization techniques that fully utilize the GPU hardware. 
On a series of experiments on the well-known QAPLIB data sets, our solutions, on average run an
order-of-magnitude faster than previous implementations and deliver up to a factor of 63 speedup 
on specific instances. The quality of the solutions produced by our implementations of \textit{2opt} and \textit{tabu} is within 1.03\% and
0.15\% of the best known values. The experimental results also provide
key insight into the performance characteristics of accelerated QAP solvers. In particular, the
results reveal that both algorithmic choice and the shape of the input data sets are key factors in
finding efficient implementations.  
\end{abstract}


\end{frontmatter}

\section{Introduction}

%
%

%
%
The Quadratic Assignment Problem (QAP) is an NP-hard combinatorial optimization problem \cite{Sahni:76,zhu.curry.marquez:10}. The objective is to assign $n$ units to $n$ locations to minimize the total cost computed as the sum of the products of {\em flows} between units and {\em distances} between locations. The \textit{flow} and \textit{distance} matrices are assumed known. 
The most common Industrial Engineering application of QAP is the design of facility layouts \cite{chiang.kouvelis:96}.  
In addition, QAP has wide applicability in many different domains including, economic modeling~\cite{Koopmans.Beckmann:57}, campus planning \cite{Dickey:72}, hospital layout \cite{Elshafei:77},  scheduling ~\cite{Geoffrion.Graves:76}, ergonomic design of electronic devices ~\cite{anstreicher.brixius.goux.linderoth:02} and processor and memory layout optimization~\cite{Wess:04,Steinberg:61}. The problem of assigning docks in a cross-docking facility is modeled also as a special case of the QAP \cite{cohen.keren:09}. 



%
%
The QAP complexity and its practical and theoretical importance, have motivated researchers, over
the years, to propose many types of algorithmic solutions. In general, instances of size $n > 30$ cannot be solved exactly in a reasonable time even on
today's high performance computing (HPC) platforms~\cite{anstreicher.brixius.goux.linderoth:02,Loiola:07,Novoa:XSEDE15}. For this reason, the body of work on QAP is dominated by
heuristic and meta-heuristic solutions. The first parallel implementations
to QAP were proposed in the early 1990s~\cite{chakrapani.skorin:93,Taillard:91}. With the emergence
of Graphical Processing Units (GPUs) as a central player in the HPC world, researchers have focused their attention to accelerator-based
solutions in recent years~\cite{zhu.curry.marquez:10,boyer.baz:13,czapinski:13}. 

%
%

The development of a QAP
implementation, that can efficiently solve a variety of problem instances have proven to be
challenging. This is because the performance of QAP implementations tend to
be highly sensitive to both the size and the {\em shape} of the input data sets. QAP instances come
in many forms. For instance, the flow matrix can be dense or sparse; symmetric or asymmetric; randomly or non-randomly distributed. An implementation may take advantage of one of these properties to quickly find a good
solution but can completely collapse for instances where that property does not hold. In
most cases, the issue is not with the specific implementation but rather the algorithm
itself. For example, it has been shown that a genetic algorithm yields very high performance on sparse
data sets but its efficiency is substantially diminished when the data set is dense~\cite{Zvi:03}. 

Finding good QAP solutions becomes further complicated on GPU platforms. Permutation-based QAP
formulations\footnote{most common formulation and the focus of this work} typically operate on the flow and distance matrices. Although the size of these matrices is not
prohibitive for GPUs, the data contained within is needed by each thread in the kernel and therefore
their access must be carefully controlled to prevent lost cycles due to synchronization. Needless to
say, QAP solutions exhibit the same properties as dense matrix computations. Therefore for best
performance both the thread and the memory hierarchy must be carefully managed to find the right
balance between occupancy and data locality~\cite{Volkov:SC08,Rashid:CSC10}. These challenges with optimizing
QAPs are compounded when we have to account not only for performance but energy efficiency as well. 

%
%


In this paper, we present high-performance GPU-accelerated implementations of a \textit{2opt} and a \textit{tabu search} algorithm for solving the QAP. We parallelized \textit{2opt} because we were interested in assessing the solution quality when using a simple heuristic for the problem. On the other hand, we parallelized a \textit{tabu search} algorithm because all \textit{tabu search} variants previously studied have reported equal or better solutions compared to other approximate methods \cite{chiang.kouvelis:96,Taillard:91,skorin:94,Eric:95}.  For
each algorithm, we extract parallelism at multiple levels, taking full advantage of the target GPU
hardware. For the \textit{tabu search} algorithm, we introduce dynamic parallelism, a novel strategy, where the number of tasks to be performed in parallel is determined based on runtime
information. We also implement several code optimizations that take advantage of specific
architectural features of the GPU. 



We conduct extensive experimentation on the Texas Advanced
Computing Center (TACC) supercomputing cluster. The experimental results on two QAPLIB datasets show
that our implementations can run an order of magnitude faster than previously proposed
strategies. This increased performance does not influence the quality of the solution. In fact, the \textit{tabu search} accuracy (proximity to best known value) of the search results is
increased. Additionally, we provide a parameterized implementation of \textit{tabu search} that exposes key algorithmic
properties for tuning. The exposed parameters include number of
neighborhoods to explore, number of parallel search instances, and number of distinct random seeds used.

To summarize, the main contributions of this paper are as follows:

\begin{list2}
\item we provide two new and efficient parallel GPU implementations for solving the QAP, an
  important problem in the area of Industrial Engineering and Operations Research. 
\item we implement several code optimizations that can be applied to other heuristic search
  algorithms. The optimizations include the novel use of GPU dynamic threads. 
\item we conduct experiments with a parameterized implementation of \textit{tabu search} that provide
  key insight as to how algorithmic properties influence the performance and the quality of the
  solution. 
\end{list2}
 

The remainder of the paper is organized as follows. Section 2 presents background on QAP formulation
and general purpose GPU computing. Section 3 discusses related work on parallel solutions to QAP using
GPU. Section 4 describes the two algorithms implemented. Section 5 describes our code optimization methods and the way some of them are included in an autotuning system. Section 6 presents the experimental results. Section 7 concludes the paper and discusses future research.       

%
%
%

\section{Background}

In this section, we provide background on the QAP formulation and on general purpose GPU computing (GPGPU).


\subsection{QAP Formulation}
\label{QAP}

Koopmans and Beckmann provide a formulation for the QAP~\cite{Koopmans.Beckmann:57}, which we
describe here using the notation adopted in this paper. Let
$n$ be the problem size and $F$ and $D$ be two given 
$n \times n$ matrices that represent \textit{flows} between units and \textit{distances} between locations, 
$F=[f_{kl}]$ and $D=[d_{ij}]$. Consider the set of positive integers $1,2,...,n$ and let $\Pi_{n}$
be the set of all permutations of $1,2,...,n$. The QAP finds a permutation $\pi^*$
$\in$ $\Pi_{n}$ such that the sum of the products in equation (1) is minimized. In (1) $\pi_i$ denotes the unit number stored in location $i$ of permutation $\pi$ 
\begin{equation}
\label{naive_cost}
z_\pi=\sum_{i=1}^{n}\sum_{j=1}^{n}f_{\pi_{i}\pi_{j}} \cdot d_{ij}
\end{equation}

Koopmans and Beckmann also re-stated the formulation above as a quadratic 0-1 integer programming
problem~\cite{Koopmans.Beckmann:57}. Since the quadratic 0-1 and the formulations above are equivalent, we omit the presentation of the 0-1 integer programming formulation. It helps to keep this paper at a reasonable length. 
In the remainder of the paper,  we will refer to Eq.~\ref{naive_cost} as the cost of a given permutation $\pi$ or the cost of a given solution.
All the solutions generated by the algorithms studied are ranked on the basis of this equation. 

\subsection{General Purpose GPU Computing (GPGPU)}


In the past, GPUs were special-purpose hardwired application
accelerators, suitable only for conventional graphics applications. Modern GPUs are fully
programmable, autonomous parallel floating-point processors which can simultaneously execute the
same program instruction on multiple data streams. Nvidia, the leading manufacturer of GPUs, released the Compute Unified Device Architecture (CUDA),
a parallel computing platform and programming model that provides a C programming language interface
to program the GPU hardware. CUDA enables dramatic increases in computing performance by harnessing
the power of the GPUs.     

One appealing characteristic of the GPU is that it efficiently launches many threads and executes them in
parallel to enable computational throughput across large amounts of data. Each thread runs the same
program named a kernel. Threads are grouped into thread blocks and all threads in a thread block may
cooperate to solve a sub problem. 
A grid is a set of blocks which are completely independent. 
A warp is a group of threads within a block that are
launched together and execute together. Warp size is typically $32$ threads on current generations of GPUs. Each thread block is mapped to
one or more warps. When the thread block size is not a multiple of the warp size, the unused threads
within the last warp are disabled automatically.  

The GPU memory is organized in a hierarchical way. \textit{Register} and \textit{Shared} memory
reside on the GPU chip. Data stored in \textit{Register} memory is visible only to the thread that
wrote it. \textit{Shared memory} is also a fast memory that can be read and written by all threads
within a block but not across blocks. \textit{Global}, \textit{Constant} and \textit{Texture} memory
reside off chip. \textit{Global memory} is accessible for read and write across blocks and also permits communication between CPU and GPU. \textit{Constant} and \textit{Texture} memories are also accessible by all threads but only for reading. The size of the \textit{Constant} memory cannot be dynamically changed.  


Many different factors affect the performance of GPU programs including efficient distribution of data
processing between CPU and GPU, the level of required communication and synchronization among
threads, the optimization of data transfer between the different parts of the memory hierarchy, and the capacity
constraints of these memories.





\section{Related Work}
\label{related}
As mentioned earlier, interest in exploiting GPU hardware for solving large QAP instances is
relatively recent and as such there are very efforts in this area. In this section, we look at GPU-accelerated solutions for QAP 
and compare our approach with previous strategies. 


\subsection{Solving QAP using GPU}
%

The previous work most related to our research is the one by Zhu {\em et al.}~\cite{zhu.curry.marquez:10}. The authors
proposed a single-instruction multiple data tabu search (SIMD-TS) for QAP using 
a single GPU on a personal computer. The parallelization consisted of running 6144
simultaneous independent tabu searches (6144 threads, 32 blocks, 192 threads per block) on 128
processors. Texture memory was used to store the distance and flow matrices. However, the authors
reported that the experimental performance was affected by the small cache size (8KB) of the texture
memory. The authors set the size of the dynamic \textit{tabu list} of smaller than the one we used
($[3,3+n/10]$  vs. $[0.1*n, 0.33*n]$) and avoided the use of slow memory by eliminating the creation
of the two-dimensional array to identify \textit{tabu} pairwise exchanges (the reader will find more
details on our {\em tabu}  algorithm in Section 4). Instead, to block the interchange for a period
of time, the authors marked as tabu one or both of the units in the selected pairwise exchange.
\emph{Zhu et al.} proposed as future research the implementation of a long term memory feature. To
assure each thread searches a different but promising area, the authors implemented random selection
of a set of 4 diversification and intensification operations every $m$ iterations and for the
experimentation $m$ was set to twice the dimension of the problem. The authors demonstrated the
implemented algorithm was effective. They used instances of different sizes ($30 < n <90$) available
at QAPLIB and the worst performance gap is 0.85\%. This percentage results from comparing the
solution they reported and the latest best known solution for the instance named \textit{tai80a}.  

Czapinski proposed an effective parallel multi-start tabu search (PMTS) for the QAP on the CUDA
platform~\cite{czapinski:13}. His technique diversifies an initial solution, runs
multiple tabu searches on each diversified solution, and re-starts the search with the best
solutions after a certain number of iterations. 
The proposed search benefits from communication between parallel \emph{tabu
search} instances by passing the best obtained solutions to the CPU. Once the CPU chooses new
configurations, the parallel tabu search is re-started in the GPU. From initial experiments the
author agreed with~\cite{zhu.curry.marquez:10} that 192 threads per block was the best
choice. Instances of size 50-70 ran faster on the GPU when compared to a six-core MPI implementation.  

Other heuristic approaches to solving QAP on the GPU have used genetic algorithms~\cite{tsutsui.fujimoto:09}
and ant colony optimization (ACO) combined with \emph{tabu search}~\cite{Tsutsuifastqap}. In regard
to exact approaches, \cite{DBLP:journals/corr/GoncalvesPDBF15} implement a level 2 Reformulation and
Linearization Technique, RLT on a heterogeneous system comprised of CPUs and GPUs.  The GPU is used
to perform the cost concentration, cost spreading, and cost transfer between complementary
coefficients operations. These operations are steps in the dual-ascent algorithm used to compute
valid and tight lower bounds for the cost of the optimal solution at each fathomed node in the
branch and bound tree. The application containing the branch-and-bound algorithm is executed in the
CPU using multiple CPU threads.The authors are able to solve exactly by the first time the instances
{\em tai35b} and {\em tai40b} proposed in \cite{Eric:95}. 

To the best of our knowledge, our work is the first to successfully parallelize the {\em tabu search}
meta-heuristic with the \textit{recency-based} feature implemented serially in
\cite{chiang.kouvelis:96}. Furthermore, we propose a novel approach of dynamic adaptation that had
previously not been applied to this particular domain.

\section{Algorithms}


\subsection{Serial 2opt}

\textit{2opt} is a heuristic method originally proposed by Croes for solving the Traveling
Salesman Problem~\cite{Croes:58}. To solve a QAP of size $n$, \textit{2opt} starts with a random
initial solution or permutation of the integers $1,2,...,n,$ stored in the 
array $\pi$. The cost of the initial solution, $z_\pi$, is computed using
Eq.~\ref{naive_cost}. The initial solution is also stored in the \textit{current solution} and in the \textit{best-solution-so-far} arrays. The cost of the initial solution is stored in the \textit{best-cost-so-far} variable. The algorithm moves forward by exploring solutions in a {\em neighborhood}. To get a single neighborhood solution, \textit{2opt} randomly selects two positions, $i,j \in \pi$ and performs a pair\-wise exchange of their content. This move is simple and convenient since it doesn't change the location of any other unit (or department if modeling a facility layout problem as a QAP) in the permutation. The simple move also permits a fast evaluation of the cost of the new solution. For a problem of size $n$ and a given current solution, the size of the neighborhood after performing all pairwise exchanges is $n*(n-1)/2$. Fig 1 illustrates the
systematic way in which all six pairwise exchanges are done for a permutation of size four.

\begin{figure}[hbtp]
\centering
\includegraphics[scale=0.25]{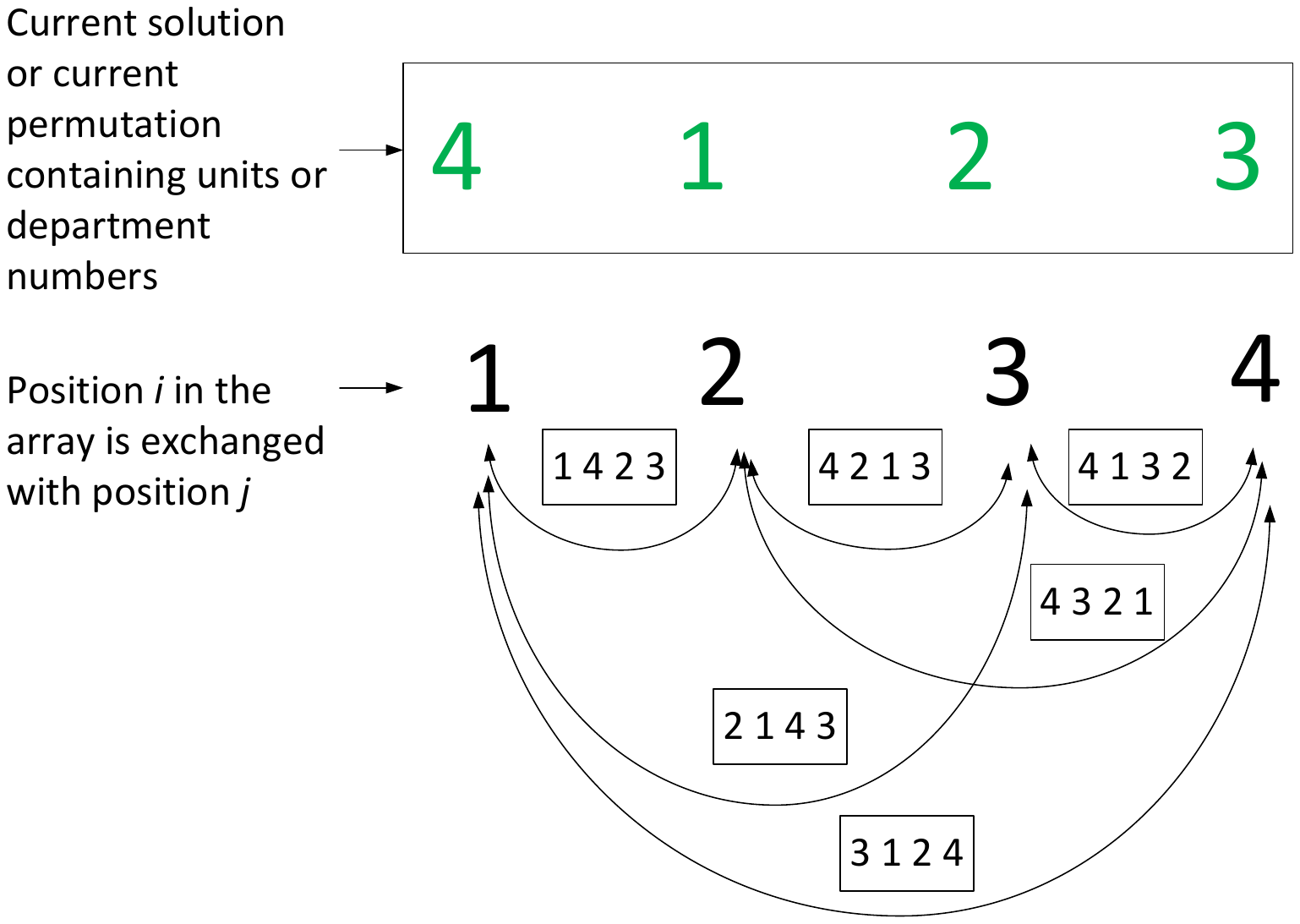}
\caption{All pairwise exchanges in a permutation of size four}
\end{figure}


We apply Eq.~\ref{new_cost} proposed by Burkard and Rendl~\cite{Burkard.Rendl:84} for computing the change  (i.e., {\em delta}) in cost after a pairwise exchange. The formula considers the case in which both flow and distance matrices are asymmetric. It can be applied to the cases of symmetric flows and/or distances without loss of generality. The formula computes the cost in linear time (i.e., $O(n)$ vs. $O(n^2)$ for formula in Eq.~\ref{naive_cost} ) 
\begin{eqnarray}
\label{new_cost}
\nonumber \Delta_{ij} &=& (d_{ji}-d_{ij})(f_{\pi_{i}\pi_{j}}-f_{\pi_{j}\pi_{i}})\\
 \nonumber &+&\sum_{k\in n\backslash\{i,j\}}((d_{jk}-d_{ik})(f_{\pi_{i}\pi_{k}}-f_{\pi_{j}\pi_{k}})\\
&+&(d_{kj}-d_{ki})(f_{\pi_{k}\pi_{i}}-f_{\pi_{k}\pi_{j}}))
\label{opt_cost}
\end{eqnarray}

The \textit{2opt} algorithm computes the costs of  all neighborhood solutions, finds the lowest cost
solution, stores it as the \textit{current solution}, and updates the\textit{ best-solution-so-far}
and its cost, if needed.  This process is repeated for a predetermined number of iterations, at
which point the best solution found and its cost is returned.   

\subsection{Parallel 2opt}



In the parallel version of \textit{2opt}, we use different random seed values to generate a set of
$N$ random permutations of the integers $1,2,...,n$ (i.e., $\Pi={\pi_1, \pi_2, ..., \pi_N}$). The
permutations (i.e., initial solutions) are stored in a matrix of size $N \times n$. Each permutation
is assigned to a single GPU thread which 
computes its cost using Eq.~\ref{naive_cost}. Fig. 1 illustrates the case in which five random
permutations (i.e., $N = 5$ ) of size $n = 4$ are assigned to five threads. Next, each thread independently
performs all pairwise exchanges on the initial solution and computes the associated costs using
Eq.~\ref{new_cost}.  






The number of iterations to perform in parallel \textit{2opt} is set as a function of
$n$.  After the number of iterations is reached, each thread returns to the CPU the
\textit{best-solution-so-far} array and the \textit{best-cost-so-far} value. The returned
\textit{best-cost-so-far} values allow the CPU to find one or several permutations of minimum cost,
output the results and terminate the algorithm.

\begin{figure*}
\centering
\includegraphics[scale=0.40]{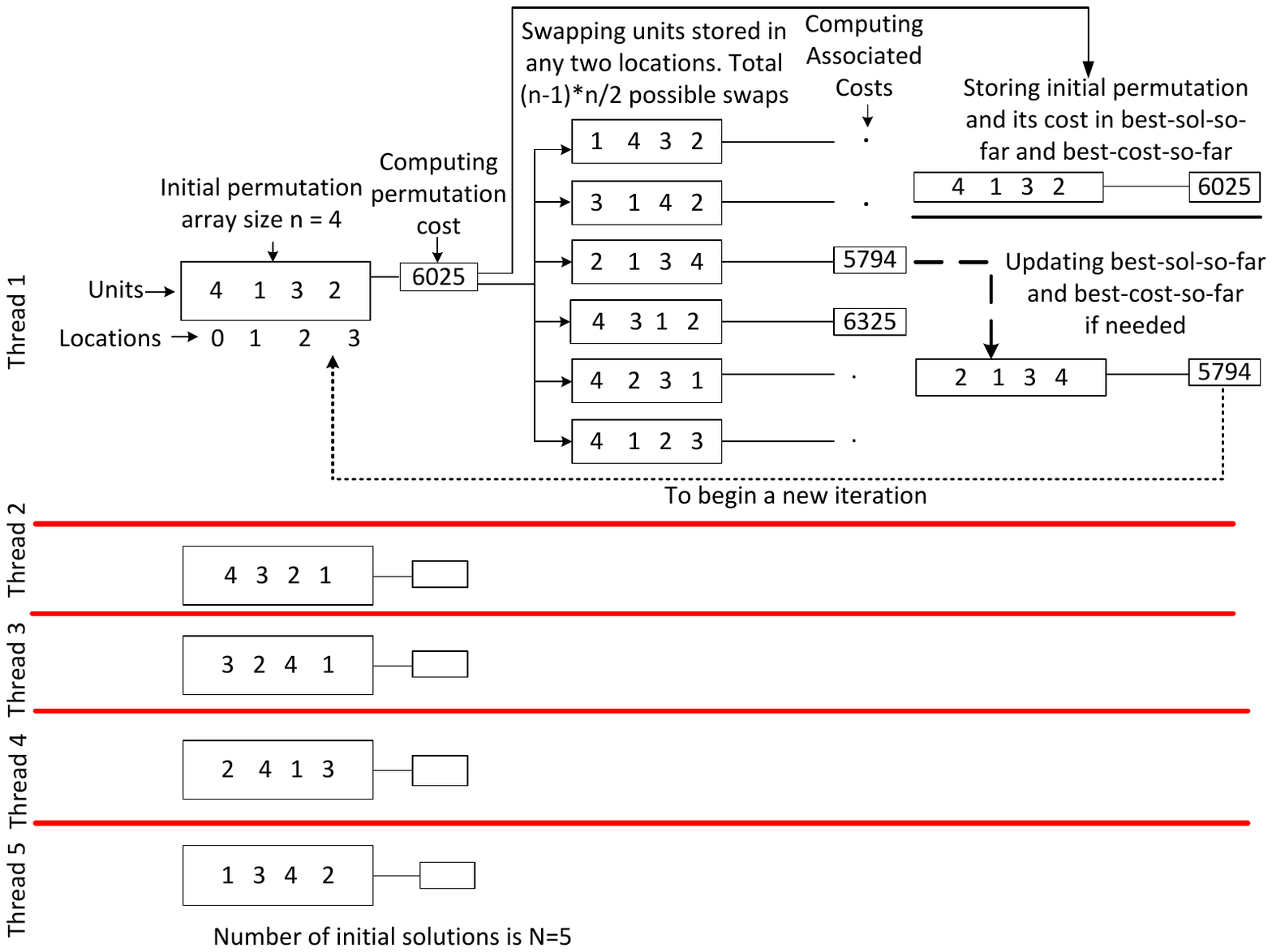}
\caption{2opt Search in GPU}
\end{figure*}

\subsection{Serial Tabu Search}
The serial \textit{tabu search (TS)} we selected to parallelize resembles the elementary
(i.e., simple) {\em TS} in
\cite{Taillard:91}. The reader not familiar with {\em TS} may consult
\cite{glover.laguna:97}.



Our elementary {\em TS} starts with a randomly generated permutation $\pi$ of the integers
$1,2,...,n$. This permutation becomes the \textit{initial solution} and
the \textit{current solution}. The algorithm stores the \textit{initial solution} and its cost
(computed using Eq.~\ref{naive_cost}) in \textit{best-solution-so-far} array and
\textit{best-cost-so-far} variable, respectively. The algorithm also sets a predetermined number of
iterations and an iteration counter to the initial value of one.  

At any iteration, neighbor solutions from a \textit{current solution} are generated from the pairwise exchange procedure described in the serial \textit{2opt} section and exemplified in Fig. 1. {\em TS} computes the costs of all neighbor solutions using Eq.~\ref{new_cost} and it chooses the neighbor solution with the lowest cost. If the solution selected is forbidden (i.e., it is in the \textit{tabu list} and does not satisfy the aspiration criteria),  {\em TS} drops this solution from consideration and proceeds to select the next lowest cost solution. The meaning of a
\textit{tabu list} and \textit{aspiration criteria} in our {\em TS} algorithm are explained in the next paragraph. The solution finally selected becomes  the \textit{current solution}; it may be a
non-improving move with respect to the previous \textit{current solution}. If the cost of the
\textit{current solution} is less than the cost of the \textit{best-solution-so-far}, \textit{best-solution-so-far} and \textit{best-cost-so-far} are updated. The iteration counter increases by one and {\em TS} goes back to repeat all the steps described in this paragraph. This iterative process follows until the iteration counter equals the predetermined iterations. 

The \textit {tabu list} stores solutions that the {\em TS} method does not want to select in
the next few iterations. The objective of the \textit{tabu list} is to avoid a cycling behavior. For
instance, if the search is in a solution that corresponds to a local minimum, the best move in the
next iteration could be a deteriorating one. If the local minimum solution is not stored in the
\textit{tabu list}, in a new iteration the algorithm will return to this previous solution and then
cycling around the local optimum will occur. Since the \textit{tabu list} may forbid critical promising
moves, our {\em TS} method includes the feature known as aspiration criteria to override the tabu status of a solution. The aspiration criteria we use
allows the algorithm to select a tabu move if it leads to a solution whose cost is better than the
cost of the  \textit{best-solution-so-far}.

Our {\em TS} algorithm implements the \textit{recency-based} memory proposed in
\cite{chiang.kouvelis:96} and the dynamic \textit{tabu list} size cited in
\cite{chiang.kouvelis:96}, \cite{Taillard:91}, and \cite{Eric:95}. Both features are explained in
the next 3 paragraphs. Authors in \cite{chiang.kouvelis:96} claimed that these features plus
intensification strategies and a long-term memory structure to further implement diversification
strategies lead {\em TS} to converge to very good solutions at a reasonable speed regardless of the
initial solution.  Motivated by the very good numerical results obtained in \cite{Taillard:91} with
elementary {\em TS} method, we opted to not use further intensification and diversification
strategies besides the dynamic \textit{tabu list} size. \cite{Taillard:91} also
mentioned that {\em TS} implemented only with a  \textit{tabu list} (i.e., just short-term memory)
has no advantage on being restarted.   

The \textit{recency-based} feature in \cite{chiang.kouvelis:96} keeps track of the number of
iterations in which a move or pairwise exchange will be tabu using an $n \times n$ matrix named
\textit{Tabuarr}. Originally all cells in \textit{Tabuarr} have zeroes. For $i$ $<$ $j$, (i.e., upper
triangle of the \textit{Tabuarr}), the \textit{i-th row}  and \textit{j-th column} identifies 
the move that results if the unit stored in the permutation $\pi$ at location $i$ is interchanged
with the unit stored at location $j$. Every time units in positions $i$ and $j$ 
are exchanged, the cell $Tabuarr(i, j)$ (for $i$ $<$ $j$) stores an integer
value equal to \textit{$current\_iter + t$} where $current\_iter$ is the current iteration number
and $t$ is a randomly generated integer that facilitates the implementation of a dynamic
\textit{tabu list} size. Thus, if $taburarr[i][j] \le current\_iter$, the exchange of units $i$ and
$j$ is not tabu.   

As suggested in \cite{glover.laguna:97}, the cells in $Tabuarr[i][j]$ (for $i$ $>$ $j$)
(i.e.,  lower triangle of \textit{Tabuarr}) may store the number of times units $i$ and $j$  have been exchanged. Thus, if at iteration one, units in
positions 2 and 3 are interchanged, the cell $Tabuarr[3][2]$ becomes 1, and if at iteration six
these units are interchanged again, the cell $Tabuarr[3][2]$ becomes
2. This frequency of use information is a long-term memory structure helpful to diversify the
search. In our {\em TS} algorithm we store these values in the lower triangle of
\textit{Tabuarr}. However, we diversify the search only through the dynamic \textit{tabu list} size.   

Taillard \cite{Taillard:91} mentions that the choice of the size of the
\textit{tabu list} is critical to diversify the search. Cycling may occur if the \textit{tabu list}
size is too small. Promising moves may be
forbidden if the list is too large. It will deviate also the exploration to solutions of lower quality and increase the number of iterations to find a good or optimal solution. 
To overcome this problem, we implement a
variable \textit{tabu list} size. Since the minimum and maximum list size is problem dependent, we
experimented with the recommendations in \cite{chiang.kouvelis:96} and \cite{Taillard:91}. For most
of the problems studied, we set the list size in the interval [$0.1n$ and $0.33n$]. At every
iteration, when a selected move needs to be set as tabu, our {\em TS} algorithm throws a random
number in the interval and stores this value in $t$.     

\subsection{Parallel Tabu Search}

The {\em TS} algorithm we implement in each GPU thread is depicted in the flowchart in Fig 3. In the
remaining sections of this paper we refer to this algorithm as {\em tabu}. A set of $N$ initial random
permutations of the integers $1,2,...,n$ is generated on the CPU and stored in a matrix of size
$N \times n$. Each GPU thread receives a  single permutation or \textit{initial solution} to execute
concurrently the same {\em TS} instructions. For the step of generating and evaluating the neighborhood of a \textit{current solution} (box number six in the first column of the flowchart), we take advantage of CUDA dynamic parallelism. Each parent thread calls $n*(n-1)/2$ child threads (CT's) to compute the cost of a single pairwise exchange using Eq.~\ref{new_cost}. 
Flow and distance matrices needed to compute the costs of any neighbor solution are maintained in global device memory to be accessible by all CT's. 

Using the information returned by the CT's, each parent thread identifies
the best cost move (i.e.  pairwise interchange) and {\em tabu} proceeds to determine if the chosen move is not tabu or if it is tabu but satisfies the aspiration criteria. 
If this is the case, the selected move becomes the \textit{current solution}; otherwise {\em tabu} identifies the next
best pairwise interchange. The step of identifying a valid pairwise exchange is
done for as many times as necessary. If there are no more pairwise exchanges to select, {\em tabu} stops prematurely. However,  an appropriate choice of the \textit{tabu list}
size will avoid this.  The number of iterations in {\em tabu} is repeated as a function of $n$. After the total number of iterations is reached, each thread returns to the CPU the \textit{best-solution-so-far}  and \textit{best-cost-so-far}. The CPU identifies the permutation(s) with the minimum cost and {\em tabu} terminates once the solution (i.e. permutation) and its cost is output to a file.

\begin{figure*}
\centering
\includegraphics[scale=0.55]{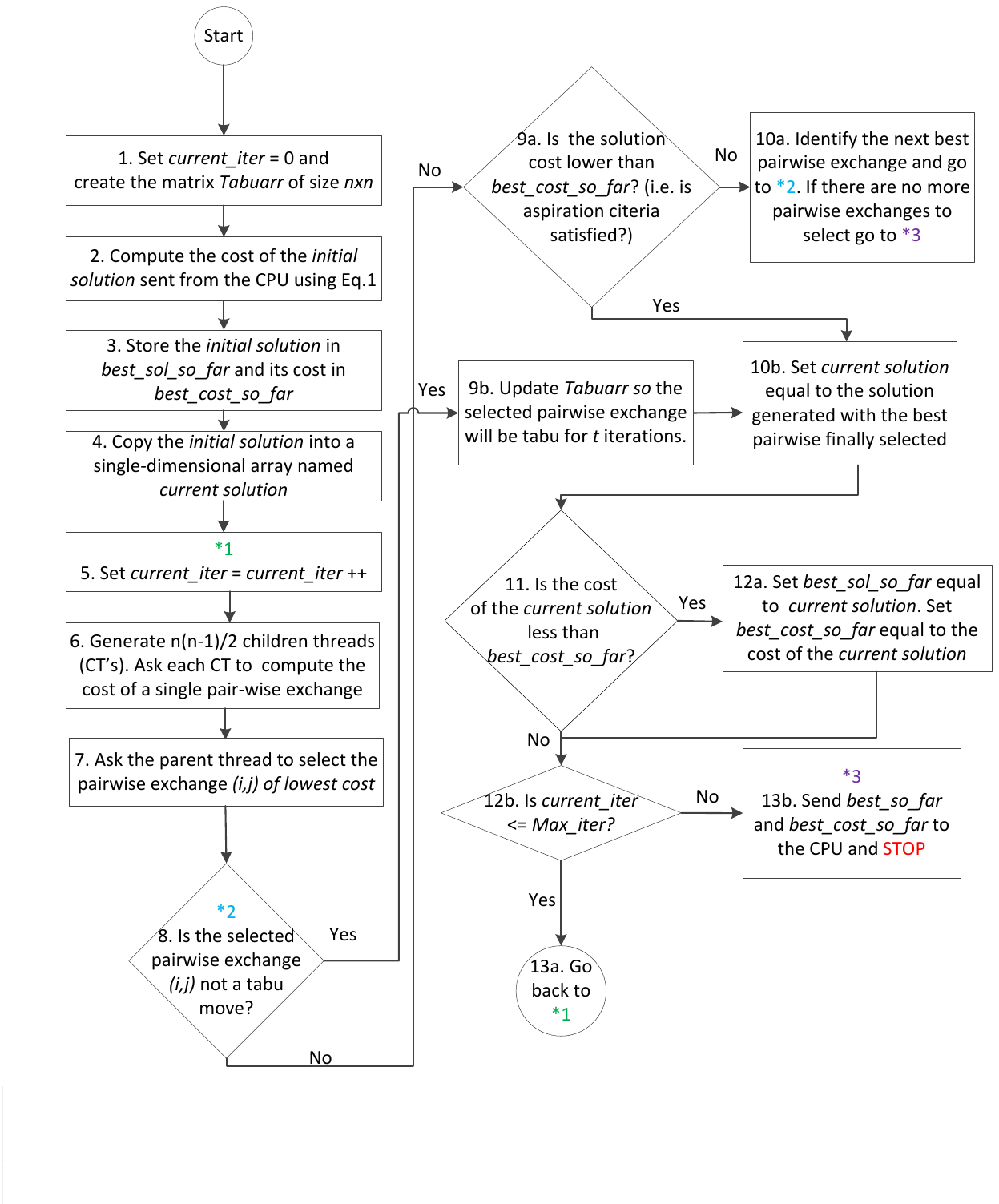}
\caption{ {\em tabu} algorithm executed at each GPU thread}
\end{figure*}


\section{Code Optimization}

We enhance our parallel implementations of {\em 2opt} and {\em tabu} in several ways to take
advantage of hardware features of the target GPU architecture. Some transformations are applied by
hand at the source-code level, while others are incorporated into an autotuning system~\cite{Sarangkar:JPCC12}
for automatic application. Below we discuss the main code transformations.  

\subsection{Shared Memory}
Effectively utilizing the shared memory hierarchy is a critical aspect of GPU performance. To
address this issue, we implement a version of {\em 2opt} that exploits inter-thread data
reuse via shared memory. In this variation, portions of the {\em flow} and {\em distance} matrices
pertinent to a single {\em neighborhood} are copied into shared memory.
The idea is to restrict memory accesses related to neighborhood exploration within the
shared memory allocated to each thread block. By avoiding the use of global memory to access the {\em flow} and {\em distance} matrices, the need for non-local memory access for
each thread is reduced, resulting in lower bandwidth requirements for the entire kernel. We 
discuss experimental results with shared memory in Section~\ref{exper:shared}.  

\subsection{Thread Configuration}

It has been shown that effectively managing the GPU thread hierarchy is instrumental in producing
high-performing GPU codes~\cite{Magni:SC13,Unkule:CC12}. On one hand, not having enough computation
per thread or per block can inhibit parallelism. On the other hand, thread coarsening or block fusion
can lead to problems with poor memory reuse. Previous implementations of QAP have all used a
fixed-size thread configuration for all instances. In this work, we develop a strategy that automatically
discovers good thread configurations. 

The QAP solvers we implemented follow a fairly simple scheme of task decomposition, where each
thread works on a separate instance of the search and all computation is done on a single
GPU {\em grid}. Nevertheless, even for this simple scheme, the choice of the number of threads per block (and
consequently, total number of blocks) can have a huge impact on performance. For this reason, we
parameterize each code variant along the thread and block dimensions.  

The number of threads and blocks are determined by the number of initial feasible solutions. In our implementations, the total number of threads across all
blocks equals $N$, the number of initial solutions. The range of the initial solutions in the search
space is determined by the quality of the solution produced. Prior studies have shown that fewer
than $2^9$ instances can impede solution quality while more than $2^{14}$ instances start to produce
diminishing returns~\cite{Chaparala:XSEDE14}.   

The number of blocks is determined by evenly dividing the total number of threads. We also ensure
that each block contains threads in multiples of {\em warp size}. If this is not the case it inevitably
leads to inefficient use of GPU resources. The maximum number of threads per block is further
constrained by the maximum number of threads allowed per block on the target platform ($2^{10}$ on Tesla
K20c). Thus, the thread configuration is a three dimensional space that can be expressed as a set as follows 

\begin{align*}
\hspace{1 in} \mathcal{T} = \{(N,t,b) & \mid 1024 \geq N \geq  12288 {\rm \ and\ } N \bmod 32 = 0,\\
            & 32 \geq t \geq 1024 {\rm \ and\ } t \bmod 32 = 0,\\
            & b = N / t \ {\rm\ where\ } N \bmod t = 0\  \}\hspace{0.8  in} (3) 
\end{align*}

where $N$ = feasible solutions, $b$ = number of blocks and $t$ = number of threads. 

During the tuning process we search values of $N$ and $t$ that meet the above constraints. The value
of $b$ is computed from $N$ and $t$. 

\subsection{Dynamic Parallelism}

The most beneficial code transformation in our implementations is the use of dynamic
parallelism. Most recent Nvidia GPUs are equipped with a feature that allows a CUDA kernel thread to
create and launch threads at runtime~\cite{NVIDIA:12}. The main advantage of dynamic parallelism is that the number
of threads to be launched does not need to be determined prior to compilation and can be adjusted
based on the size and shape of the input data set and other runtime values. In
\cite{dimarco.taufer:13}, the authors quantified the performance gains of dynamic parallelism in two
clustering algorithms.  

We take advantage of dynamic parallelism in the parallel implementation of {\em tabu}. In our
algorithm, neighborhoods are created through pair-wise exchanges of two locations. For each point in
the neighborhood the cost function needs to be evaluated. We make the observation that (i) the cost
functions can be computed independently and (ii) the number of cost functions to be evaluated
depends on the neighborhood size which in turn depends on the size of the input data set. The second observation
implies that we cannot create the parallel threads for neighborhood exploration at compile
time. This makes the cost computation tasks ideally suited for dynamic parallelism. 
 In our implementation, each parent thread, which represents one instance of a tabu search, launches $k$
child threads to explore a neighborhood in parallel. The value of $k$ is determined based on the
size of the input. For a problem size of $n$, the value of $k$ is $(n - 1) * n / 2$. Thus, the
size of a neighborhood that is explored in parallel grows quadratically with the size of the
input. This rate of growth makes intuitive sense since the growth in the overall space is
exponential with respect to the input size.

\section{Experimental Results}
\label{section:exper}

\subsection{Experimental Setup}

\subsubsection{Platforms}
The computational experiments were primarily executed on the Stampede cluster at TACC.
Stampede is a 10 PFLOPS Linux cluster based on 6,400+ Dell Zeus PowerEdge server nodes,
each outfitted with 2 Intel Xeon 8-Core 64-bit E5 processors (2.7 GHz) and an Intel Xeon Phi
Co-processor (1.1.GHz). Each node runs CentOS 6.3 (2.6 32x86\_64 Linux kernel). The nodes are
managed with batch services through SLURM 2.4. Stampede has 128 compute nodes outfitted with a 
single Nvidia K20 GPU on each node with 5GB of on-board GDDR5 memory. Each K20 GPU has 2496 CUDA 
cores distributed over 13 streaming multiprocessors (SM's). Each SM can hold a maximum of 2048
thread contexts 
The clock speed for each core is 0.706 GHz, L1 cache size is 64 KB/SM and L2 cache size is 768
KB (shared).  

For comparison purposes, serial CPU and OpenMP variants of the \textit{2opt} code were developed and compiled with GCC
Version 4.4.7. The CUDA code for the parallel \textit{2opt} and \textit{tabu} algorithms were compiled with {\tt nvcc} using CUDA version 5.5. The {\tt sbatch} script
was used to submit jobs to the cluster and to specify the node configuration. We
ran four jobs simultaneously by assigning each job to a different Stampede node. This significantly
expedited the experimentation phase.   

In addition to Stampede, we also ran experiments on a local
server with a six-core Intel Sandybridge processor. This server is equipped with a Tesla K20c NVIDIA
GPU which has the same configuration as the GPUs on Stampede. 
This server runs Ubuntu 12.04. 

\subsubsection{Data Sets}
To evaluate our implemented algorithms, we used datasets from QAPLIB, a library of published test problems
for QAP described in~\cite{burkard.karish.rendl:91}.  

{\em Lipa} instances come from problem generators described in \cite{li.pardalos:92}. These generators
provide asymmetric instances (i.e. non-symmetric flow and/or distance matrices) with known optimal
solutions. The {\em Taixxy} datasets are proposed in \cite{Taillard:91}. Instances named {\em
  Taixxa} are uniformly generated, {\em Taixxb} are asymmetric and randomly generated and {\em
  Taixxc} occur in the generation of grey patterns. Other problem sets are introduced in
\cite{Eric:95}. Each implemented algorithm was executed eight times on each given instance. The
execution times and the accuracy metrics reported in this paper are explained in detail in the
next subsection.  



\begin{table*}[t]
\centering
\caption{Execution time and accuracy comparison of {\em 2opt} and {\em tabu}}
\scriptsize{
\begin{tabular}{lllllll}\\\hline
& \multicolumn{2}{c}{Zhu et. al.} & \multicolumn{2}{c}{\em 2opt}	& \multicolumn{2}{c}{\em tabu}\\	
Problem	 & accy. & time(s) &	accy.	& time(s) &	accy. &	time(s)\\\hline
tai30a & 0.00 &  18.60  & 1.10 & 3.84 & 0.00 & 2.57 \\
tai30b & 0.00 &  192.00  & 0.00 & 3.78 & 0.00 & 3.01 \\
tai35a & 0.00 &  309.60  & 1.77 & 7.03 & 0.00 & 4.7 \\
tai35b & 0.00 &  331.20  & 0.01 & 6.90 & 0.00 & 10.39 \\
tai40a & 0.07 &  442.20  & 1.55 & 11.83 & 0.00 & 16.46 \\
tai40b & 0.00 &  508.20  & 0.02 & 11.68 & 0.00 & 9.23 \\
tai50a & 0.58 &  1,210.80  & 1.78 & 29.40 & 0.24 & 119.87 \\
tai50b & 0.05 &  574.20  & 0.15 & 29.17 & 0.00 & 63.88 \\
tai60a & 0.45 &  1,144.80  & 2.50 & 62.15 & 0.28 & 328.02 \\
tai60b & 0.12 &  2,091.00  & 0.23 & 61.19 & 0.09 & 239.08 \\
tai80a & 0.73 &  11,230.20  & 2.35 & 202.11 & 0.55 & 773 \\
tai80b & 0.25 &  10,976.40  & 0.52 & 199.20 & 0.17 & 541.87 \\
tai100a & 0.72 &  23,215.80  & 2.35 & 501.65 & 0.60 & 1912 \\
tai100b & 0.53 &  33,167.40  & 0.89 & 493.62 & 0.41 & 86.58 \\
lipa70a & 0.00 &  1,172.40  & 0.77 & 117.08 & 0.00 & 171.18 \\
lipa90a & 0.00 &  7,585.20  & 0.64 & 327.19 & 0.00 & 995.29 \\\hline
mean & {\bf 0.22} & & {\bf 1.03} & & {\bf 0.15} & \\\hline  
\end{tabular}}
\label{primary}
\end{table*}



\subsection{Performance and Accuracy}

Table~\ref{primary} reports the performance and accuracy of the fully accelerated versions of {\em
  2opt} and {\em tabu}.  For all implementations, the number of initial random solutions generated for each instance was $N=6144$. The accuracy
score is computed using the following formula 
$$\hspace{0.1 in} accy~=~ (best\_cost\_from\_8\_runs\ -\ best\_known\_cost)/best\_known\_cost \hspace{0.2 in}(4)$$
Thus, lower value implies better accuracy and a score of $0.00$ means the search was able to discover
the best known cost for a specific instance. The best known cost for a  particular instance is derived 
from previously published results~\cite{burkard.karish.rendl:91}. As reference, we compare the
performance of the two algorithms with previously published results from Zhu {\em et al} \cite{zhu.curry.marquez:10}. We refer to the
Zhu version of the algorithm as {\em zhu} in the rest of this section. Performance is reported as kernel execution time in seconds.


We observe that in terms of performance, {\em 2opt} yields the best results, achieving on average, a
factor of  $33.28$ and $1.75$ better execution times over {\em zhu} and {\em tabu}, respectively. However, {\em
  2opt} does suffer somewhat from lower accuracy. On average, {\em 2opt} has an accuracy score of
$1.03$, which is higher than both {\em zhu} and {\em tabu}. In terms of cost and performance, {\em
  tabu} provides the best results. Not only does it achieve an impressive factor of $18.96$ speedup over
{\em zhu}, it also handily beats previous versions in terms of accuracy. The average accuracy score
for {\em tabu} is $0.15$, a substantial improvement over {\em zhu} and {\em 2opt}. In 9 of the 16
instances {\em tabu} is able to discover the best known solution. On all 16 instances it is able to
find a better solution than both {\em zhu} and {\em 2opt}.  

\subsection{Comparison with Parallel CPU Implementation}

\begin{figure}
\centering
\subfigure[performance] {
  \includegraphics[scale=0.35]{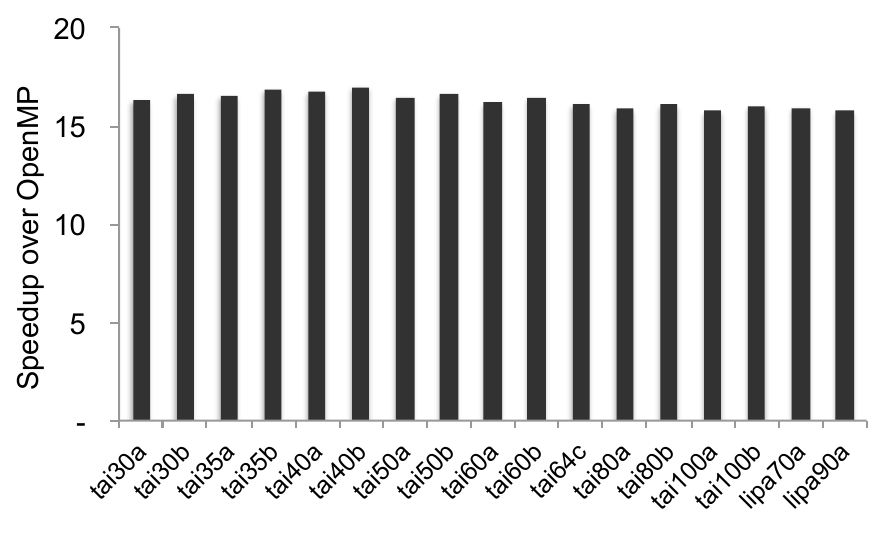}
}
\subfigure[\% of proximity to the best known solution] {
\includegraphics[scale=0.35]{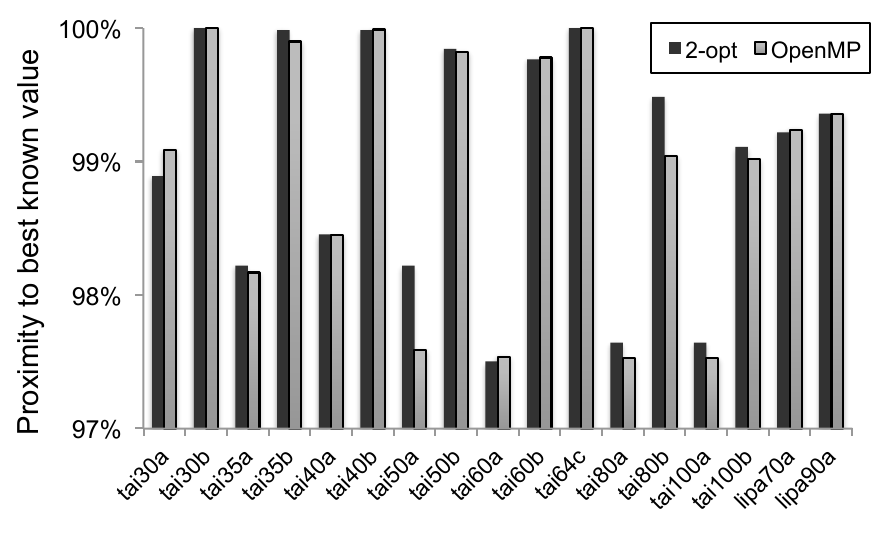}
}
\caption{{\em 2opt} and {\em OpenMP} comparison}
\label{comp_openmp}
\end{figure}

Fig.~\ref{comp_openmp} compares the performance and the percentage of {\em proximity to the best
  known solution} for {\em 2opt} and the OpenMP version of {\em 2opt} (i.e., {\em OpenMP}). {\em
  2opt} achieves at least a factor of 16 speedup over {\em OpenMP} 
on all problem sizes. We attribute this performance gains mainly to the additional computation power
available on the GPU. {\em OpenMP} was implemented using 16 threads which proved to be optimal for the compute
node configurations on the computation cluster. On the other hand, {\em 2opt} was designed to make use
of {\em all} available SMs on the target GPU allowing it to achieve more parallelism on different problem
instances. In terms of proximity to the best known solution, there is no clear advantage for either {\em OpenMP} or {\em
  2opt}.  On some instances {\em 2opt} is significantly better while on others {\em OpenMP}
yields a better solution. This is an expected result as both versions employ a random heuristic for searching.

\subsection{Thread Block Configuration}

\begin{figure}[t]
\centering
\subfigure[lipa70] {
\includegraphics[scale=0.35]{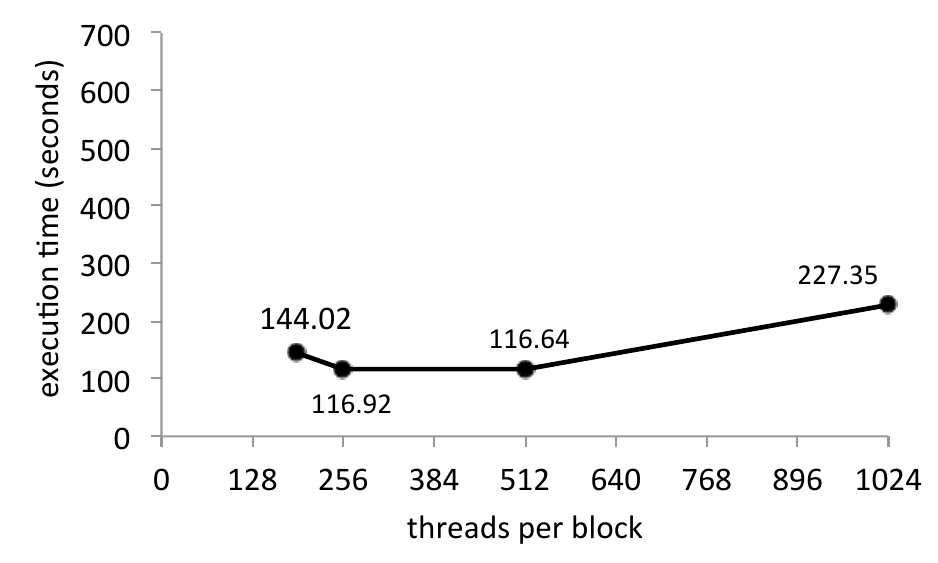}
}
\subfigure[lipa90] {
\includegraphics[scale=0.35]{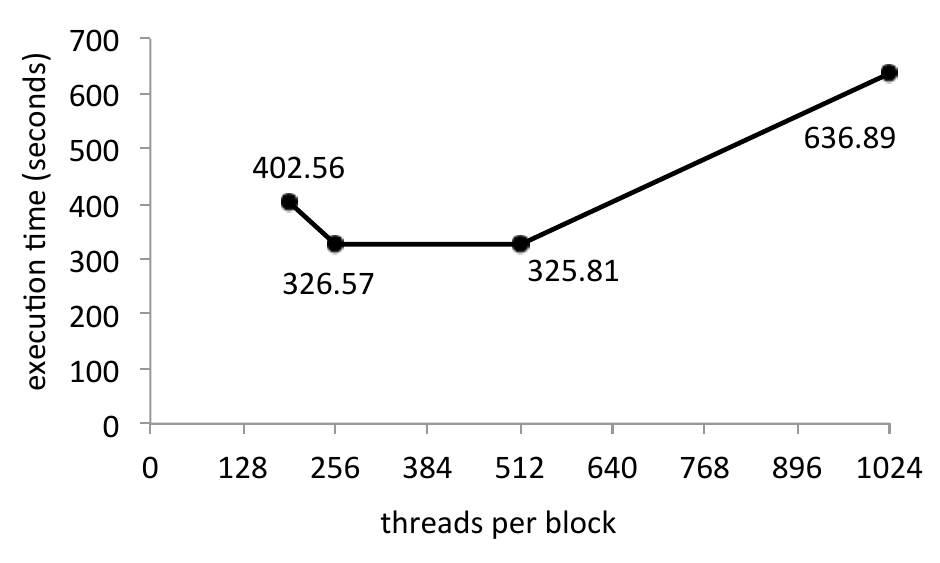}
}
\subfigure[tai100a] {
\includegraphics[scale=0.35]{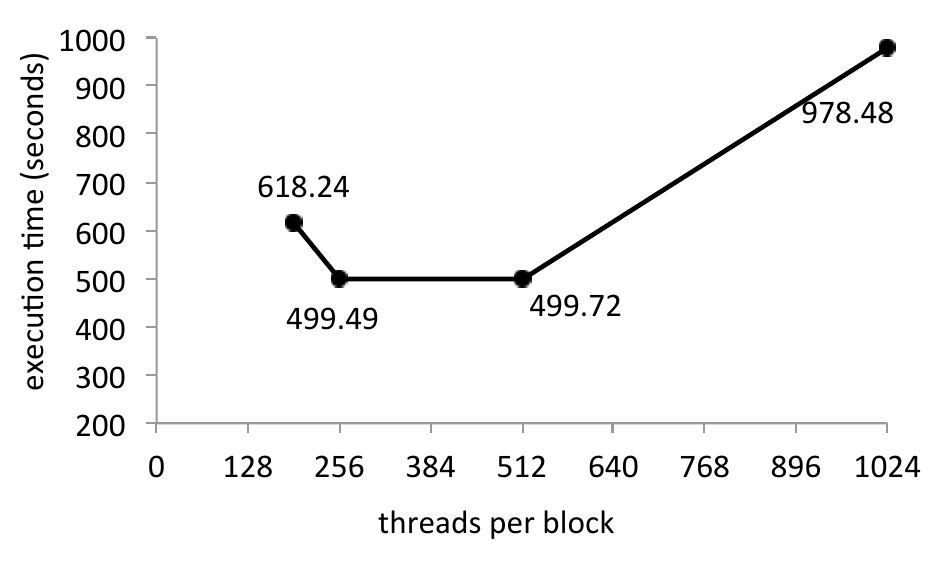}
}
\subfigure[tai100b] {
\includegraphics[scale=0.35]{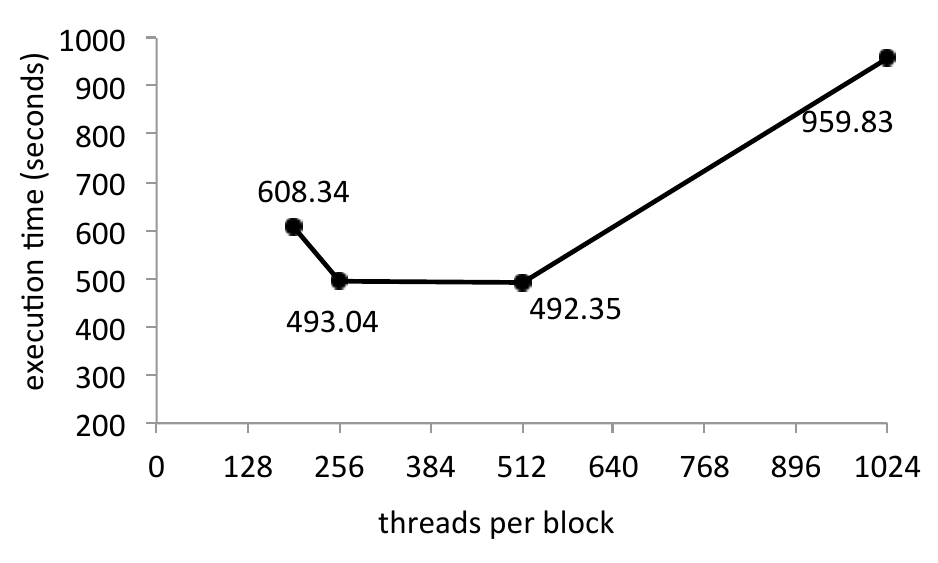}
}
\caption{Execution time variation in {\em 2opt} for varying thread configurations}
\label{thread_config}
\end{figure}

We ran a series of experiments to find a suitable thread configuration for {\em 2opt}. We
parameterized the algorithm and executed the code with different thread and block parameters to vary
the number of active warps per SM and attain different levels of
occupancy. Fig.~\ref{thread_config} presents selected results from these
experiments. The figures reveal that the best performance for {\em 2opt} is not necessarily
achieved at maximum threads per block, in spite of the fewer synchronization events occurring
in those implementations. For both {\em Lipa} and {\em Taillard} data sets, the highest performance
is achieved at 256 threads per block. We attribute this performance gain to better register usage
and shared memory utilization. These results corroborate results from earlier studies on GPU
occupancy and data locality~\cite{Volkov:SC08,Unkule:CC12}.

\begin{figure}[t]
\centering
\includegraphics[scale=0.50]{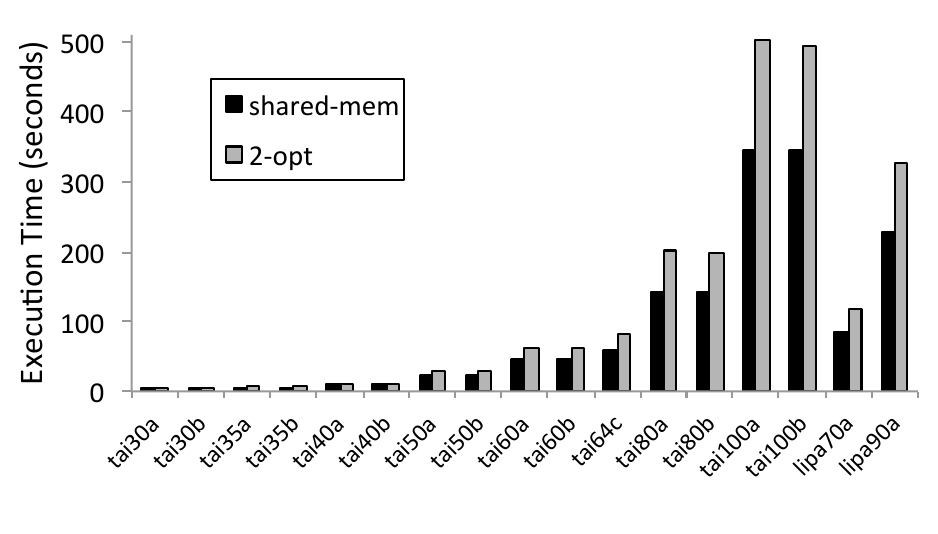}
\caption{{\em 2opt} performance with shared memory allocation}
\label{shared}
\end{figure}

\subsection{Shared Memory}
\label{exper:shared}
To optimize memory access, two key data structures, {\em flow} and {\em distance}, were allocated
to shared memory. Fig.~\ref{shared} shows performance results for the implementation of {\em 2opt}
with shared memory allocation. We notice that the shared memory implementation provides yet more
performance improvements over the highly efficient non-shared memory version of {\em 2opt}. These gains stem
from two different sources. First, allocation into shared memory replaces many of the global memory
accesses with accesses to shared memory that posses lower latencies. Second, because each thread in a block
accesses the data structures in every iteration, the shared memory allocation helps exploit the abundant inter-thread data
locality exhibited by these threads.




\subsection{Tabu Algorithmic Parameters}

To better understand how different parameters of {\em tabu} affect the accuracy and performance under
various data sets, we developed a paremeterized version of the code. The following parameters were
exposed to an external tuning system

\begin{list2}
\item number of neighborhoods (i.e. number of iterations)
\item search instances (i.e., number of parallel searches launched)
\item random seeds (i.e., number of times search is repeated with a new and distinct random number seed)
\end{list2}

In this subsection, we explore the sensitivity of {\em tabu} to these parameters. For these
experiments, we present data from {\em tai40a} and {\em tai40b}. 

\begin{figure}
\centering
\subfigure[quality of solution] {
\includegraphics[scale=0.40]{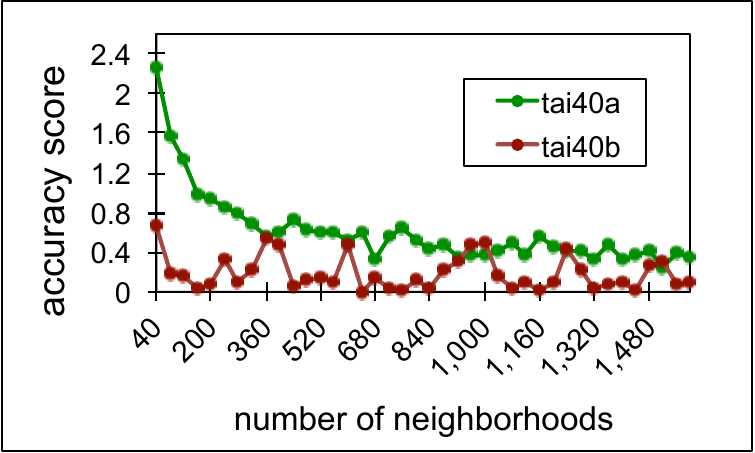}
}
\subfigure[execution time] {
\includegraphics[scale=0.40]{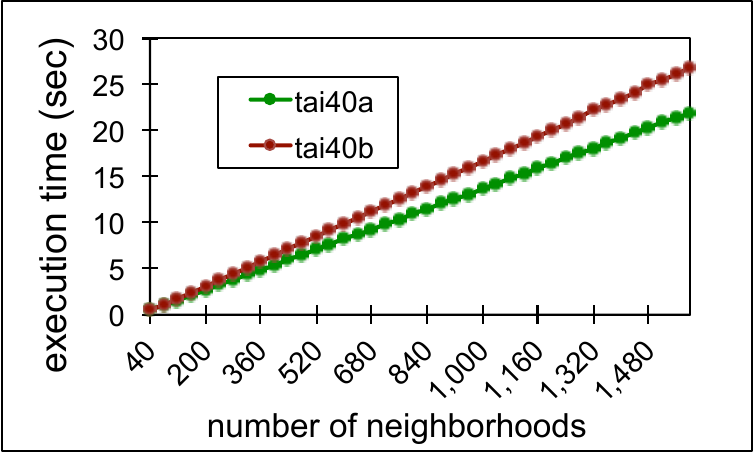}
}
\caption{{\em tabu} sensitivity to number of neighborhoods}
\label{hoods}
\end{figure}

Fig.~\ref{hoods} shows the variations in accuracy and execution time  of {\em tabu} as the number of
neighborhoods is progressively increased. We observe that the number of neighborhoods explored has a
direct linear relationship with the execution time. This, of course, is intuitive. The more
neighborhoods explored the longer the execution. The accuracy numbers paint a slightly different
picture. For {\em tai40a}, the accuracy improves sharply until the number of neighborhoods reaches
400.
Beyond that, increasing the number of neighborhoods tend to have diminishing returns. For {\em
  tai40b}, the effects are more random and there is no clear evidence that increasing the number
of neighborhoods improves the quality of solution. 

\begin{figure}
\centering
\subfigure[quality of solution] {
\includegraphics[scale=0.40]{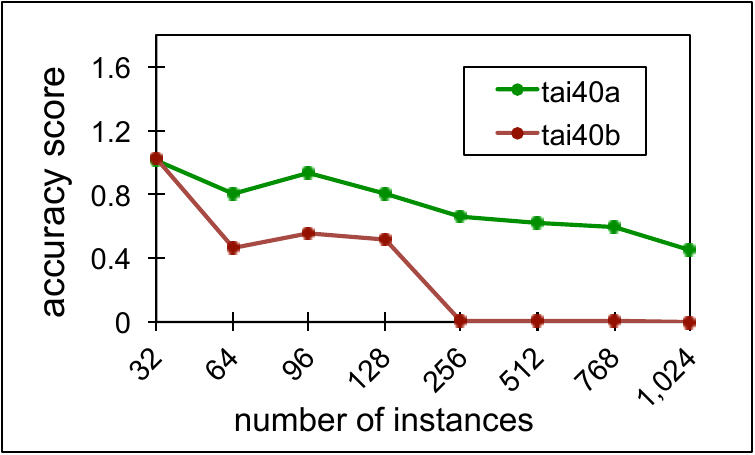}
}
\subfigure[execution time] {
\includegraphics[scale=0.40]{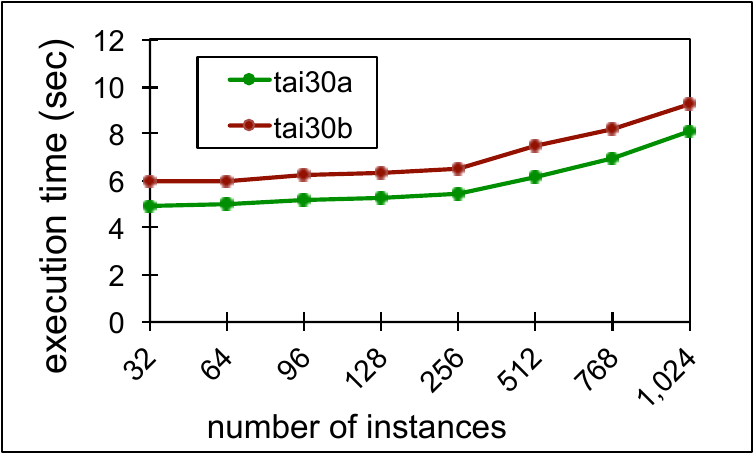}
}
\caption{{\em tabu} sensitivity to number of search instances}
\label{insts}
\end{figure} 

\begin{figure}
\centering
\subfigure[quality of solution] {
\includegraphics[scale=0.40]{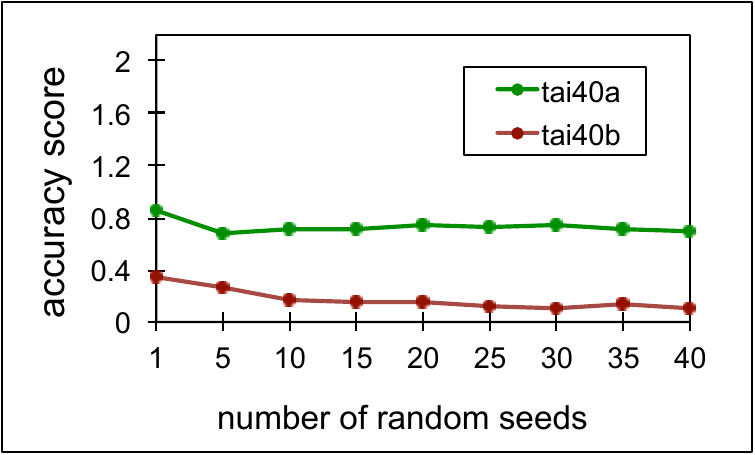}
}
\subfigure[execution time] {
\includegraphics[scale=0.40]{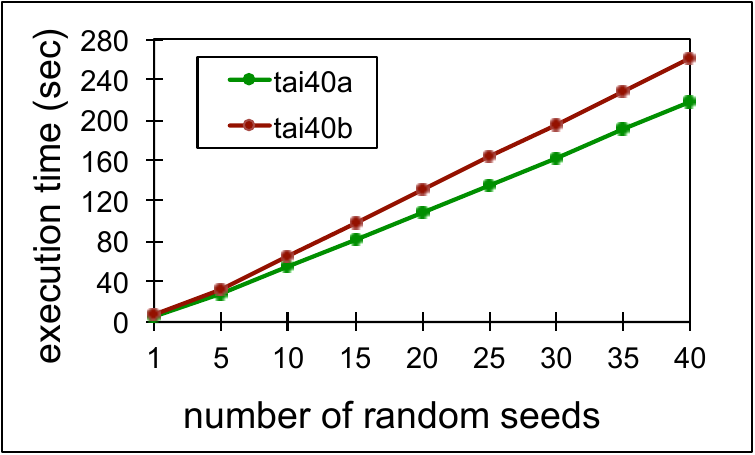}
}
\caption{{\em tabu} sensitivity to number of random seeds}
\label{restarts}
\end{figure} 

Fig.~\ref{insts} shows the sensitivity of {\em tabu} as the number of search instances is
varied. The number of instances maps to the number of threads that can be launched on one
multiprocessor (SM) on the GPU. For this reason, only powers-of-two values are chosen. The maximum
number of instances is bounded at 1024 by the physical capacity of the device. We observe that
number of instances has little effect on the execution time. This is understandable, since all
instances work in parallel. The slight increase in execution time that we observe comes from the added
overhead of thread creation. In terms of accuracy, we start to see diminishing returns for both {\em
  tai40a} and {\em tai40b} beyond 256 instances. Thus, for these data sets, 256 search instances
appears to be the ideal choice. 

Finally, we look at the impact of number of random seeds. Fig~\ref{restarts} shows the accuracy and
execution time variations as a function of number of times the search is repeated with a different
random seed. The accuracy score reported is the minimum found if using $k$ different random seeds, averaged over 8 runs. We observe that random seeds have little impact on {\em tai40a} and {\em tai40b}. For execution time, again there is a
linear relationship, as each repetition requires new instances of the search to be launched.

\section{Conclusions and Future Work}
\label{section:conclusions}

\label{conclusions}
This paper presented two GPU-accelerated solutions to the Quadratic Assignment Problem. The
implemented {\em tabu} algorithm is very efficient and accurate. Its average accuracy is 0.15\% on
the instances studied. The implemented {\em 2opt} algorithm has a better performance but it  less
accurate.  On the experiments performed, its average accuracy was 1.03\%. The {\em tabu} algorithm
exploits the CUDA dynamic parallelism available in the Nvidia K20 GPU 
card. We conclude that GPU and dynamic parallelism are attractive tools to use in implementation of heuristic
search algorithms.  



The accessibility to the Stampede cluster reduced significantly the time to complete the
experimentation phase. The on-line documentation from TACC and the suggestions from its staff
members were very helpful. It should motivate more Industrial Engineering and Operations Research
practitioners towards the use of a computational cyberinfrastructure similar to the Stampede
cluster.

 
An amenable way for exploiting the Stampede supercomputer features is to develop an MPI/OpenMP
implementation. Although a CPU-based implementation will allow us to scale to
larger datasets, we speculate that this will not lead to significant increase in performance. The task
granularity is fairly small and is more suitable for mapping to a GPU. 

We plan to incorporate a long-term \textit{frequency-based} memory feature that uses the
information currently stored in the lower diagonal of the \textit{Tabuarr} matrix. For some
instances, this feature could diversify the search even more and could find solutions that may beat
the best known ones. Furthermore, we plan to investigate the implications of nested parallelism in
the tabu implementation. In this approach, the child threads will dynamically invoke new
threads to extract more parallelism during neighborhood exploration. 



\section{Acknowledgments}
The authors acknowledge the Texas Advanced Computing Center (TACC) at The University of Texas at
Austin for providing high performance computing resources that have contributed to the research
results reported within this paper. URL: http://www.tacc.utexas.edu.  The second author 
acknowledges support from the National Science Foundation through awards CNS-1253292 and
CNS-1305302. 

\section{References}

\bibliographystyle{elsarticle-num} 
\bibliography{gpu} 



\end{document}